\documentclass
[aps,prc,amsmath,amssymb,floatfix,nofootinbib]
{revtex4}

\usepackage[dvips]{graphicx}         %
\usepackage{bm}                      
\usepackage{mathptmx}                
\usepackage{dcolumn}                 
\usepackage{multirow}                %

\def\bea{\begin{eqnarray}} \def\eea{\end{eqnarray}}
\def\beq{\begin{equation}} \def\eeq{\end{equation}}
\def\bal#1\eal{\begin{align}#1\end{align}}
\def\bse#1\ese{\begin{subequations}#1\end{subequations}}

\begin{document}

\title{Proton-proton $^1S_0$ pairing in neutron stars}
\author{
Wenmei Guo$^{a,b}$, J.M. Dong $^{c}$,X. Shang $^{c}$, H.F. Zhang$^{d}$, W. Zuo$^{c}$\\
M. Colonna$^{b}$,U. Lombardo$^{b}$ \footnote{Corresponding author
lombardo@lns.infn.it at: Laboratori Nazionali del Sud (INFN), Via
S. Sofia 62, 95123 Catania, Italy, phone: +39\ 095\ 542\ 277, fax:
+39\ 095\ 71\ 41\ 815}} \affiliation{$^{a}$ Institute of
Theoretical Physics, Shanxi University, 030006 Taiyuan, China\\
$^{b}$ Laboratori Nazionali
del  Sud (INFN), Via S. Sofia 62, 95123 Catania, Italy \\
$^{c}$ Institute of Modern Physics (CAS), Lanzhou 730000, People's
Republic of China\\$^{d}$School of Nuclear Science and Technology,
Lanzhou University, Lanzhou 730000,People's Republic of China}

\date{\today}

\begin{abstract}
The onset of $^1S_0$ proton spin-singlet pairing in neutron-star
matter is studied in the framework of the BCS theory including
medium polarization effects. The strong three-body coupling of the
diproton pairs with the dense neutron environment and the
self-energy effects severely reduce the gap magnitude, so to
reshape the scenario of the proton superfluid phase inside the
star. The vertex corrections due to the medium polarization are
attractive in all isospin-asymmetry range at low density and tend
to favor the pairing in that channel. However quantitative
estimates of their effect on the energy gap do not give
significant changes. Implications of the new scenario on the role
of pairing in neutron-star cooling is briefly discussed.
\end{abstract}



\maketitle

\section{Introduction}
The interest for the superfluidity in nuclear matter started soon
after the pairing in nuclei\cite{cooper}, but it has been rising
after the neutron-star physics had addressed various superfluid
states\cite{alpar}: $^1S_0$ neutron-neutron pairing in the crust
made of a low-density neutron gas in equilibrium with the ion
lattice, $^1S_0$ proton-proton pairing in the outer and inner
core, where the proton density is kept low enough by the
$\beta$-equilibrium with neutrons, electrons and muons, and
finally $^3PF_2$ neutron-neutron pairing deep inside the core.
Superfluidity in the crust is a key ingredient for the
understanding of many different phenomena in compact star physics,
from the cooling of new born stars \cite{latt,burr}, to the
afterburst relaxation in X-ray transients \cite{red}, as well as
in the understanding of glitches\cite{fat}. In particular, the
proton pairing could play an important role in the cooling of the
NS core either within the models of standard cooling or in the
case of minimal cooling \cite{page}. Therefore, due to the lack of
observational evidence, it is required a good theoretical
description of the pairing in NS. Recent calculations
\cite{dong,dong1}, in fact, have shown that the $^3PF_2$ pairing
is quenched down by medium polarization effects, independently of
the experimental and theoretical uncertainties of the high-density
interaction. As to $^1S_0$ proton-proton pairing it is already
well known \cite{zuo,zuo1} that the repulsive interaction via
three-body force between diproton  pairs with the neutron
background cannot be neglected, and in fact it shrinks the pairing
domain to a significant extent: according to Ref.~\cite{zuo} the
upper baryonic density limit is $\rho=0.3 fm^{-3}$. At such  a
density the proton fraction (in $\beta$ equilibrium with neutrons
and electrons) is quite close to the threshold of direct URCA
(dURCA) processes~\cite{durca}. The consequence is that the
$^1S_0$ proton pairing could not attenuate the dURCA rapid cooling
rates, as required by observations. An additional correction to
the pairing interaction is due to the medium polarization in the
framework of the induced interaction theory \cite{babu,sjob,back}.
The latter was first found to reduce spin-singlet pairing in pure
neutron matter\cite{clark}, later confirmed by other
calculations\cite{ains,schul}, but in symmetric nuclear matter,
the opposite event happens\cite{cao}. After the extension of the
induced interaction theory to asymmetric nuclear matter
\cite{zhang}, it is now possible to include the induced
interaction in the spin-singlet proton-proton pairing in neutron
stars. In Ref.~\cite{solov} it is shown that the induced
interaction is attractive in the spin-singlet channel at low
density, that once more could reshape the role of proton-proton
pairing in neutron stars. To which extent this happens is the main
scope of the present note.

\section{Proton superfluid state in the $^1S_0$ channel}\label{pp}
 In the study of the
proton-proton (pp) pairing in neutron stars one should consider
that the protons are in equilibrium with neutrons and leptons,
that makes the proton fraction quite low and the neutron fraction
quite large dependent on the distance from the center of the star.
This situation permits the proton superfluid to exist in
equilibrium with non superfluid neutron background up to high
baryonic densities. However the contribution of the three-body
force $V_{ppn}$ to the pairing interaction cannot be neglected,
 because the neutron gas interacting with protons is quite dense
\cite{zuo}. An additional peculiarity of the proton pairing in
neutron stars is that the self-energy and vertex corrections due
to the medium polarization are isospin dependent. Including all
many body effects  the gap equation takes the form
\begin{eqnarray}
\Delta(k)&=& -\frac{1}{\pi}\int_0^\infty  k'^2{\rm d} k' \frac{Z^2
V(k,k')}{E_{k'}} \Delta(k') \\
V(k,k') &=& V_2(k,k')+V_3(k,k')+V_{ind}(k,k'),
\end{eqnarray}
where $E_k=\sqrt{(\epsilon(k)-\epsilon_F
)^{2}+\Delta_{k}^{2}}$,$\epsilon_F$ being the Fermi energy. $V_2$
is the realistic pp interaction in the spin-singlet state, $V_3$
the effective 2BF, obtained from 3BF $V_{ppn}$ saturating the
neutron states suitably weighted on the correlation with the two
protons \cite{average}. The self-energy corrections affect both
the single-particle energy $\epsilon(k)-\epsilon_F
=(k^2-k_F^2)/2m^*$ via the effective mass $m^*$, and the depletion
of the Fermi surface via the quasi-particle strength Z. The
chemical potential $\mu$ is self-consistently calculated with the
gap from the conservation of the quasi-particle number. The
self-energy effects have been tested in many works, resulting
always into a reduction of the energy gap. Finally, the various
particle-hole collective excitations of the medium entail vertex
corrections $V_ind$ which strongly depend on the isospin
composition of the system \cite{zhang}. The effect of those
corrections embodied in $V_{ind}$ on the proton-proton
spin-singlet pairing is the main scope of the present
investigation.

\begin{figure}[htbp]
\begin{center}
\includegraphics[width=0.75\textwidth]{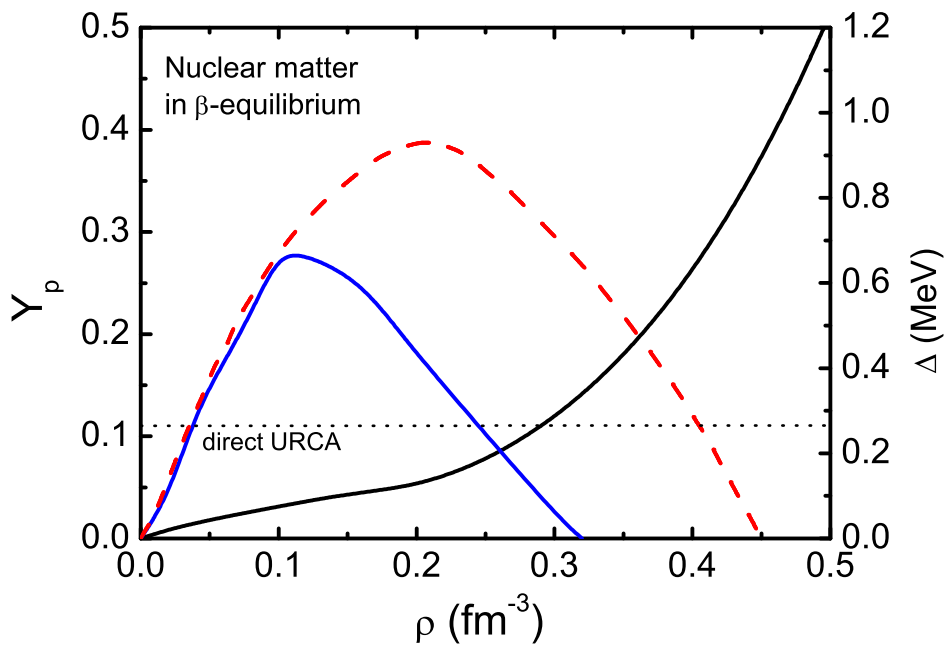}
\caption{Proton fraction vs. total density in $\beta$-equilibrium
nuclear matter (left y-axis). The dotted horizontal line marks the
proton threshold  of dURCA processes. In the same figure the
$^1S_0$ proton gaps (right y-axis) with two-body (dashed curve)
and two- and three-body forces (solid line) without screening
effects.}\label{fig:NS}
\end{center}
\end{figure}
Being 3BF repulsive, its effect in the $^1S_0$ proton channel is
that of reducing the gap magnitude but also the density range of
superfluidity in NS \cite{zuo}. The numerical results are reported
in Fig.1, where gaps with only 2BF and 2BF plus 3BF are plotted.
They have been obtained employing as pairing interaction Argonne
V18 2BF \cite{av18} and consistent meson-exchange 3BF
\cite{average}. The self-energy effects have been neglected
because first we want to focus on the interplay between 3BF and
induced interaction. As above mentioned, the figure shows that,
when including 3BF, the overlap between the region of $^1S_0$
proton pairing and the region of dURCA processes disappears, that
implies that proton-proton pairing cannot attenuate its strong
cooling effect.

\section{Particle-particle induced interaction in the proton spin-singlet channel}

\begin{figure}[htbp]
\begin{center}
\includegraphics[width=0.5\textwidth]{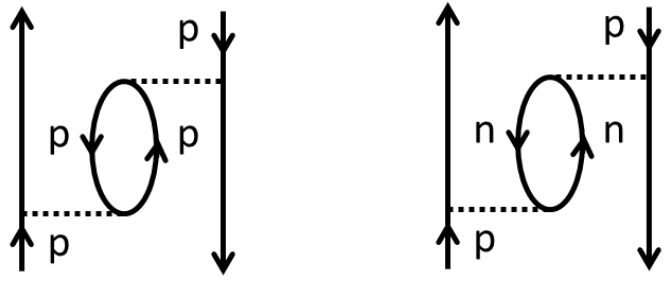}
\caption{Diagrams of the particle-hole  proton and neutron medium
particle-hole excitations participating in the induced
interaction.The vertex insertions represent the full particle-hole
interaction itself.}\label{fig:dia}
\end{center}
\end{figure}

 The vertex corrections to the $^1S_0$
proton-proton pairing interaction have been determined in the
framework of the induced interaction theory \cite{babu,sjob,back}.
This theory combines the effect of the medium polarization on the
ph effective interaction with the effect of the latter on the
medium polarization itself. This main feature entails the removal
of the instability of nuclear matter due to the low-density
density fluctuations ($F_0 < -1$).  The induced interaction theory
has been applied to nuclear pairing in the extreme situations of
symmetric nuclear matter (SNM) and pure neutron matter (PNM)
\cite{cao}: More recently it has been extended to asymmetric
nuclear matter \cite{zhang} mainly for application to neutron-star
physics. Calculating the p-h induced interaction
$({\cal{J}}^{p-h}_i)^{S_{ph}}_{\tau,\tau'}$ in asymmetric nuclear
matter the p-p spin-singlet (S=0) and spin-triplet (S=1)
components are derived as linear combinations of the $S_{ph}$
terms (see Ref.~\cite{zhang} for details). The corresponding
diagrams are plotted in Fig.2. In the case of proton-proton
spin-singlet (as the $^1S_0$ two-particle state) it was found that
\begin{equation}
({\cal{J}}^{p-p}_i)^{S=0}_{p,p} = \frac{1}{2}[
({\cal{J}}^{p-h}_i)^0_{p,p}\,-3\,({\cal{J}}^{p-h}_i)^1_{p,p}]
\label{eq:singlet}.
\end{equation}
One can see that the induced interaction is driven by the
interplay between density fluctuations ($S=0$) and spin density
fluctuations ($S=1$). The multiplicity of the latter is expected
to play, roughly speaking, the major role, considering that the
two contributions are of the same order of magnitude
\cite{heisel}. The results are shown in Fig.3 as a function of
asymmetry for a number of relevant total baryonic densities.
\begin{figure}[htbp]
\begin{center}
\includegraphics[width=0.5\textwidth]{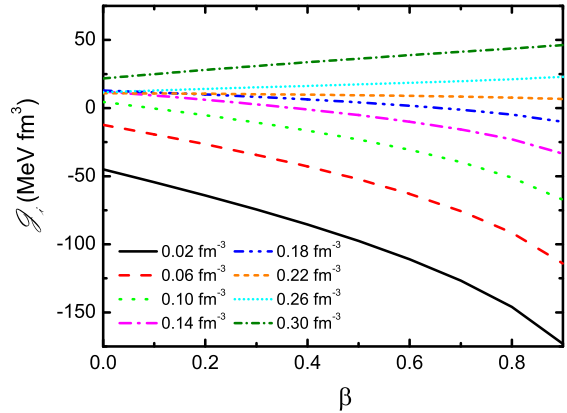}
\caption{Proton-proton induced interaction vs. asymmetry parameter
(N-Z)/A for a set of baryon densities.}\label{fig:ind}
\end{center}
\end{figure}

 A typical example of spin-singlet is
that of two-particle channel $^1S_0$, and in fact its effect has
been studied for the case of neutron-neutron pairing in that state
for the two extreme conditions of SNM and PNM \cite{cao}. It was
shown that the induced interaction is attractive in SNM and it is
repulsive in PNM. In $^1S_0$ proton-proton pairing the opposite
effect occurs, as shown in Fig.3. In this figure it is shown that
the induced interaction is almost always attractive in PNM. In
particular, it is always attractive in the full asymmetry range
from SNM to PNM for total density $\rho\le 0.1 fm^{-3}$, always
repulsive for $\rho\ge 2.2 fm^{-3}$ and first repulsive and then
attractive for intermediate densities.

\section{Proton $^1S_0$ pairing with screening}

\begin{figure}[htbp]
\begin{center}
\includegraphics[width=0.5\textwidth]{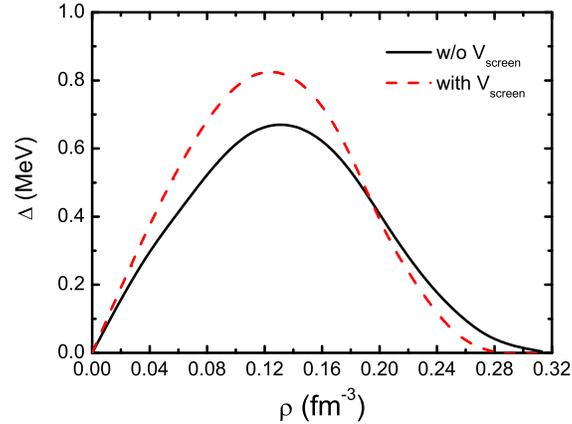}
\caption{Energy gap in $^1S_0$ proton-proton channel with and
without induced interaction.}\label{fig:gap}
\end{center}
\end{figure}

The procedure of solving the BCS equation with screening requires
a special care. In fact the induced interaction in the Landau
limit does not provide off-diagonal matrix elements of the
interaction in momentum space. However, since the collective
excitations giving rise to the medium polarization are
concentrated around the Fermi surface, the momentum transitions
from the Fermi momentum to lower and higher momenta beyond the
Fermi surface are negligible . Therefore, we can include the
screening effects as corrections to the exact solution of the gap
equation without screening in the following way. The transitions
due to the 2BF and 3BF beyond the Fermi surface can be
incorporated in the bare pairing force following a well known
procedure based on the splitting of the total momentum space into
two subspaces (see Ref.~\cite{mig} for details). Introducing small
window w around the Fermi momentum $k\approx k_F$ the gap equation
splits into two coupled equations
\begin{eqnarray}
\Delta(k_F)&=& -\frac{1}{\pi}\int_{w_<}  k'^2{\rm d} k'
\frac{1}{E_{k'}}
\tilde V(k_F,k')\Delta(k') \\
\tilde V(k,k') &=& V(k,k')-\frac{1}{\pi}\int_{w_>}  k''^2{\rm d}
k'' \frac{V(k,k'')\tilde V(k'',k')}{E_{k''}},
\end{eqnarray}
with $w_<$ ($w_>$) is the momentum interval inside (outside) the
window. It is understood that the interaction embodies $Z^2$. The
solution of the two coupled equations, Eqs.(3-4), is the same as
that of Eq.(1), independently of the window size. Since the
screening interaction is small it is reasonable to truncate to the
first order Eq.(4) in the expansion around the pairing interaction
$\tilde V^{(0)}$ without screening, so that $\tilde V = \tilde
V^{(0)}+ V_{ind}$, being $V_{ind}=V_{ind}(k_F,k_F)$. Therefore the
calculation starts from solving the exact gap equation for the
pairing interaction without screening to get $\tilde V^{(0)}$ and
then it proceeds solving the linearized gap equation with
screening, Eq.(3). Following that procedure the gap equation was
solved with the choice of Av18 2BF \cite{av18} and consistent
meson-exchange 3BF \cite{average} as pairing force and the induced
interaction described in Sec.III as screening force. The resulting
gaps are reported in Fig.3. As expected, the screening effect
produces a sensitive enhancement, over $20\%$, of the gap
magnitude due to the low density attractive effect of the induced
interaction and a small damping in the tail of the gap curve due
to repulsive effect of the high density induced interaction. It is
worthwhile comparing our results with Ref.\cite{baldo} despite the
different approaches adopted to calculate the vertex corrections.
In that paper the the latter are obtained from an RPA calculation,
that is not consistent with the Landau parameters describing the
effective interaction. A close similarity is found between our
proton gaps with those they obtain with the isoscalar Landau
parameter $F_0=-0.4$.
 A calculation including the self-energy
effect has also been performed. The effective mass and the
quasi-particle strength have been extracted from the self-energy
$\Sigma(k,E)$ calculated to the third-order of G-matrix in a BHF
code. Some results are reported in Table I. The self-energy
effects reduce sizeably the pairing gap, a bit less in the
presence of the induced interaction. This result should be taken
with some caution because a consistent calculation including on
the same footing vertex and self-energy corrections due to the
coupling to collective modes is still on the way.

\begin{table}
\renewcommand{\arraystretch}{1.5}
\begin{ruledtabular}
\vspace{1mm}
\begin{tabular}{cccccccc}
$\rho (fm^{-3})$ & $k^p_F(fm^{-1})$ &$m*/m$& $Z$ &$\tilde V_0$ & $\Delta_0(MeV)$ &$\tilde V$&$\Delta(MeV)$\\
\hline
0.08 &0.385 & 0.90 & 0.59 &-0.936   &0.0157 &  -1.03 & 0.027  \\
0.10 &0.420 & 0.85 & 0.64 &-0.808  &0.0205 &  -0.89 & 0.036 \\
0.12 &0.475 & 0.82 &0.65 & -0.723  &0.0165 &  -0.78 & 0.027\\
\end{tabular}
\end{ruledtabular}
\caption{Proton $^1S_0$ pairing in $\beta$-stable nuclear matter
in the presence of self-energy effects. $\rho$ is the total
density and $k^p_F$ the Fermi momentum of the corresponding proton
fraction. $\tilde V_0$ ($\tilde V$) the effective interaction
without (with) induced term in units of $N(0)$ and
$\Delta_0$($\Delta$) the corresponding gap. The self-energy
parameters are the effective mass $m*$ and the quasi-particle
strength $Z$.}
\end{table}

In conclusion, a theoretical treatment of the proton $^1S_0$
pairing including medium polarization in the pairing interaction
do not significantly changes the previous results obtained with
2BF and 3BF. In particular, the pairing range remains below the
activation threshold of dURCA processes, so that the proton
pairing cannot quench down the efficiency of dURCA in the NS
cooling. Moreover the strong self-energy effects make the gap
magnitude rather small on nuclear scale. Nonetheless they could
affect the long-era cooling of NS, because the thermodynamic
parameters are exponentially dependending on $\Delta/kT$, where
$T$ is the temperature.

\begin{acknowledgments}
This work was supported by INFN post-doc fellowship program
and the National Natural Science Foundation of China under
Grants No.~11705109 and 11775276.
\end{acknowledgments}

\end{document}